\documentclass[showpacs,preprintnumbers,superscriptaddress,amsmath,amssymb,floatfix]{revtex4}

\usepackage[usenames]{color}
\usepackage{epsfig,epstopdf}
\usepackage{caption}
\usepackage{amssymb}
\usepackage{graphicx}
\usepackage{subfigure}
\usepackage{float}
\usepackage{multirow}
\usepackage{booktabs}

\newcommand{\be}{\begin{equation}}
\newcommand{\ee}{\end{equation}}
\newcommand{\bee}{\begin{eqnarray}}
\newcommand{\eee}{\end{eqnarray}}

\definecolor{grey}{rgb}{0.5,0.5,0.5}

\definecolor{black}{rgb}{0,0,0}

\def \irbaddress{Rudjer Bo\v{s}kovi\'{c} Institute, Bijeni\v{c}ka cesta 54, P.O. Box 180, 10002 Zagreb, Croatia}
\def \Tuzlaaddress{Univerzity of Tuzla, Faculty of Scinece, Univerzitetska 4, 75000
 Tuzla, Bosna i Hercegovina}

\begin{document}

\title{Poles, the only true resonant-state signals,  are extracted from a worldwide collection of partial wave amplitudes using only one, well controlled pole-extraction 
method}

\author{M. Had\v{z}imehmedovi\'{c} }
\affiliation{\Tuzlaaddress \\}

\author{S. Ceci}
\affiliation{\irbaddress \\
*E-mail: alfred.svarc@irb.hr}

\author{A. \v{S}varc*}
\affiliation{\irbaddress \\
*E-mail: alfred.svarc@irb.hr}

\author{ H. Osmanovi\'{c} }
\affiliation{\Tuzlaaddress \\}

\author{J. Stahov }
\affiliation{\Tuzlaaddress \\}

\date{\today}

\begin{abstract}
 Each and every energy dependent partial--wave analysis is parameterizing the pole positions in a procedure defined by the way how the continuous energy dependence is 
implemented. These pole positions are, henceforth, inherently model dependent.  To reduce this model dependence, we use only one, coupled--channel, unitary, fully analytic 
method based on the isobar approximation to extract the pole positions from the each available member of the worldwide collection of partial wave amplitudes which are 
understood  as nothing more but a good energy dependent representation of genuine experimental numbers assembled in a form of partial--wave data.  In that way, the model 
dependence related  to the different assumptions on the analytic form of the partial--wave amplitudes is avoided,  and the true confidence limit for the existence of a 
particular resonant state, at least in one model,  is established. The way how the method works, and first results are demonstrated for the S$_{11}$ partial wave. 

\end{abstract}


\date{\today}

\maketitle

\section{Introduction}

When resonances are associated with the eigenstates of the complete Hamiltonian for which there are only asymptotically outgoing waves,  their identification  with scattering 
theory poles is unquestionable. This statement is in details elucidated in ref. \cite{Dal70}.   Consequently, in order to get the full information about physical systems and 
resonant states under observation,  we must be entirely focused onto analyzing and interpreting the scattering matrix singularities of the Mandelstam analytic function 
\cite{Man58} obtained from experiments.  While the value of the scattering amplitude on the positive energy cut defines the physical amplitude in the s or u channel depending 
whether we approach the physical axes from above or below, the simple poles which are situated on the physical axes in a subthreshold region are related to the bound states.  
As it is believed that there is no fundamental difference between a bound state and a resonance, other than the matter of stability,  when simple poles of the coupled channel 
amplitude occur on unphysical sheets in the complex energy plane, they are to be associated with resonant states \cite{Mar70}.  \\

The fact that we are trying to  extract the value of a quantity lying in the complex energy plane while performing experiments only on the physical axes, is the essence of all 
problems, and origin of many misunderstandings  occurring in the literature. Namely, each pole is not only squatting in an experimentally inapproachable domain, but is 
simultaneously governing each and every process between all allowed few body channels. However, we usually measure observables only in one channel at a time. If the 
single-channel observables are measured, we obtain the \emph{single-channel scattering amplitude}, and we only get the pole positions in one channel. Nevertheless, due to the 
Mandelstam hypothesis, these poles are affecting all channels, so we have to treat them all and not just the measured one.  Consequently,  the underlying theory which is to be 
used to find the scattering matrix amplitude must be a coupled-channel one, and of course analytic and unitary. And this is not the end.  Once we have found the coupled-channel 
scattering matrix amplitude, we have to \emph{find} and \emph{quantify} all its poles.  Unfortunately,  this is not a simple task, so each partial wave analyses, even in a 
multi-channel case, has its own way of parameterizing this inaccessible quantity. The result is that the model dependencies are introduced. \\

This brings us to various ways how the complex energy plane poles are up to the present moment parameterized in the literature. First attempts are done with single-channel 
partial wave amplitudes, and the oldest and most frequently met way is the concept of Breit-Wigner parameters. \\

The initial attempts to use the Breit-Wigner function with constant parameters to represent the scattering matrix amplitudes on the physical axes immediately revealed the fact 
that this function is too simple. More terms were needed. One had to introduce the energy dependent background, and  one had to do it in a unitary way. Unfortunately, for quite 
some time it has been known, \emph{but not commonly accepted}, that a unitary addition of background terms \emph{influences} the peak position of the scattering matrix absolute 
value on the real axes in spite of the fact that the pole position \emph{is not changed}. Peak position is an interplay of Breit-Wigner parameters and background terms. And 
\emph{the peak position} is the quantity which is usually extracted from experiments. Consequently, when Breit-Wigner parameters defined in such a manner are chosen to 
represent the pole position, they must be background dependent,  and the \emph{only case} when the Breit-Wigner parameters \emph{do exactly} correspond to the pole position is 
when we have \emph{accidently} guessed the correct form of the energy dependent background. If the background is wrong, Breit-Wigner parameters are not the pole parameters, but 
something else. And that is the reason why Breit-Wigner terms in general \emph{are not} the pole positions, and are inherently model dependent. \\

There are basically two ways to account for the background contributions. The first one, to unitary add energy dependent background terms to the constant-parameter Breit-Wigner 
function, is described afore. The second one is to allow the Breit-Wigner parameters to become energy dependent. That is predominantly done by modeling the Breit-Wigner 
width~\cite{Arndt1,Manley1,Manley2,Cutcosky1,Bat1,TPVrana}. \\

There is a number of ways to introduce energy dependent Breit-Wigner width.  In ref. \cite{Arndt1} energy dependent width is a part of resonant term of  theoretical function 
which is associated with the T-matrix near the resonance. In refs. \cite{Manley1,Manley2} energy dependent width is related to the resonant part of the S-matrix. In a method 
proposed in  ref \cite{Cutcosky1} and applied in refs. \cite{Bat1,TPVrana} width is deffined from the function consisting of a background term and Breit-Wigner shape term.  \\

One well known method for treating the nearby channel openings is the Flatt\'{e} formula. The Flatt\'{e}'s method, introduced in 1976 \cite{Fla76}, is recognizing the fact that 
the partial-wave T-matrix feels the presence of new channel openings, and it is taking it into account effectively. Flatt\'{e} proposes to modify the traditional Breit-Wigner 
form by assuming that the width with becomes proportional to the phase space. The amplitude poles are then again represented as the singularities of the modified Breit-Wigner 
function. \\

The fact that the Breit-Wigner terms in general \emph{are not} the pole positions, and are inherently model dependent, was timidly mentioned by several authors (see for 
instance ref. \cite{Pole_vs_BW}). That was first strongly pointed out by H\"{o}hler in refs. \cite{KH80,PDG1998}, where  the definition of ``local Breit-Wigner fit" and the 
concept of ``searching for the pole position" using speed plot technique were introduced.  H\"{o}hler clearly distinguished between Breit-Wigner parameters (which should  in 
the absence of a better way be obtained by locally fitting partial wave amplitudes with a Breit-Wigner function plus some background terms) and pole parameters which should be 
obtained, as he recommended , by the speed plot technique. He has always been pointing out that Breit-Wigner parameters are model dependent, and he continuously objected to 
compare them directly. His last warning was published not so long ago \cite{PDG2000}.  However, due to unclear historical reasons, the practice of direct comparing Breit-Wigner 
parameters coming from different origin continued in  Particle Data Group (PDG) compilations.  Breit-Wigner parameters, extracted with different background parameterizations 
are still directly compared \cite{PDG2010}, averages are made and error analysis is performed neglecting the fact that they may be in fact completely    \emph{differently} 
defined parameters. This practice should be abolished. \\

There is a long history of efforts to avoid the concept of Breit-Wigner parameters, and to look directly for the genuine pole positions. \\

The first, and most frequently met method, is the speed-plot technique introduced by H\"{o}hler \cite{KH80} for the single channel scattering amplitudes. It is based on the 
idea already mentioned in ref.~\cite{Mar70} that the pole position should be found by expanding the scattering amplitude in the vicinity of the pole, and the speed-plot 
technique is recommending to retain the first term only. This method is in principle acceptable if we are dealing with isolated poles far away from any nearby thresholds, but 
my fail otherwise. There is a number of cases where the methods can not be applied at all, and the best example was inability to use it to obtain the well known S$_{11}$(1535) 
resonance. The limitations of the method have been discussed by Ceci et al. \cite{Ceci08} where it has been shown that speed-plot technique is only the $N$=1 term of a more 
general but demanding ``regularization" method based on finding the N-th derivative of the scattering amplitude, and using it in a local, three-parameter fit to the partial 
wave data \cite{Mainz2010}. \\

In the early fifties the time delay  technique is introduced into scattering theory by several authors \cite{Eisenbud1, Wigner1, Wigner2, Bohm}, in a way that they obtained 
expression for the time delay in a collision. Time delay, or in another words the time lapse between asymptotic states, can be directly related with phase shift of the T 
matrix. For further details on interrelation between speed plot and time delay see ref.~\cite{Suzuki1}.  \\

The N/D method is a technique in which the dispersion relations are used to construct the amplitudes in the physical region using the knowledge of the left-hand cut 
singularities. The idea is to represent the partial-wave amplitude as a ratio of two functions. The numerator is represented with a function N(s) which is analytic in the 
s-plane  only on the left-hand cut, and the function D(s) that is analytic on the right-hand cut only.  The poles of the scattering amplitude are identified with the zeroes of 
the D(s), and the problem of extra zeroes is often difficult to be solved. The method has been introduced a long time ago by \cite{Chew60}, and since then it has been mostly 
used in meson physics, typically for cases when the knowledge about the left-hand cut is available \cite{Olle99,Ani06}. \\

All enlisted methods are good for the pole search within certain approximations, but yet we have to point out that the proper procedure to look for the scattering matrix poles 
is the full analytic continuation of scattering matrix amplitudes into the complex energy plane within a given model. \\

In coupled-channel calculations the importance of the pole-search has recently been fully recognized.  Some gro\-ups have offered more or less detailed concepts of their 
analytic continuation procedures \cite{Arn04,EBAC}, while others have reported that the complexity of the analytic continuation of all Feynman amplitudes of their model is 
beyond their reach \cite{Feu98}. Therefore, they had to rely on speed-plot technique entirely. In most cases, the analytic continuation procedure is rather cumbersome. \\

The VPI/GWU collaboration clearly distinguishes the difference between Breit-Wigner parameters and pole positions, and states that \emph{"Poles and zeros have been found by 
continuing into the complex energy plane"}. Unfortunately, they fail to provide any details of their procedure. The EBAC collaboration  also makes an analytic extrapolation of 
their amplitudes, and has recently presented a more detailed elaboration of their procedures  \cite{EBAC}. Other groups have extracted their  pole positions using 
single-channel techniques such as speed-plot and time delay \cite{Che03,Che07,Feu98,Giessen}. Recognizing the importance of a direct analytic extrapolation, Dubna-Mainz-Taipei 
collaboration has recently performed the full analytic continuation, and in ref.~\cite{Mainz2010} offered the reliable pole positions of their model.  \\

 In spite of all these efforts, the question of systematic uncertainties still remains unanswered, because each model has its own, particular analytic form.  So we wonder how 
stable, with respect to the analytic continuation procedure, these pole positions actually are.  \\

In order to get a reliable answer to this questions, we have decided to use \emph{only one method} to extract pole positions from \emph{all published partial waves analyses}, 
and compare the results. And we have chosen the T-matrix Carnagie-Melon-Berkeley (CMB) model. In other words, we take all sets of partial-wave amplitudes, treat them as nothing 
else but a good, energy dependent representations of all analyzed experimental data, and extract the poles which are required by the CMB method. In this manner, all errors due 
to different analytic continuations of different models are avoided, and the only remaining error is the precision of CMB method itself. We shall also compare the obtained 
poles with the poles of each individual publication, and draw certain conclusions about features of individual methods as well.  \\
    
The general idea of this article is to recommend the possibility how to, in a maximally model independent way, simultaneously find  all scattering matrix poles from the 
world-wide collection of partial wave amplitudes. We present the way of eliminating most systematic errors in analytic extrapolation  by using only one, well defined procedure 
to extract pole positions for published partial wave amplitudes, understanding them as nothing more but a very confident energy dependent representation of all experimental 
data.  \\
  
To avoid congesting the reader with unnecessary information, we in this paper in details illustrate how this method works for the S$_{11}$ partial wave only. We show that 
N(1535) and N(1650) S$_{11}$ resonant states are unambiguously seen in all analyzed PWA data, while the performed pole-search procedure strongly suggest the existence of at 
least one more pole position in the vicinity of 1800 MeV.  Therefore, all published PWA are consistent with the new  S$_{11}$(1846) state seen in  photo-production channel 
\cite{Che03,Yan03}. We demonstrate that the existence of fourth  S$_{11}$ state around 2100 Mev is not excluded by any PWA, and is actually favored for the hadronic 
Dubna-Mainz-Taipei amplitudes \cite{Che03,Che07}. We compare the obtained results with the results published in literature, and make a final conclusion on the actual position 
of partial wave poles. \\

However, the issue also arises how strongly the recommended pole-extraction procedure (CMB model) depends upon its own the model assumptions. Namely, CMB model has a number of 
assumptions, and it is very important to know how stable the pole positions are if CMB model choices are strongly modified. We have tested this problem extensively, and for the 
answer to this question we refer the reader to a parallel publication submitted to this journal \cite{Hedim2010}.  

\section{Formalism}
The CMB model is  isobar, coupled-channel, analytic, and unitary  model, where the T matrix in a given channel is assumed 
to be a sum over the contributions from a number of intermediate particles (resonance and background contributions). The coupling of the channel asymptotic states to these
 intermediate particles determines the imaginary part of the channel function, and is represented effectively with a separable function. 
 The real part of the channel function is calculated by the dispersion relation technique, thus ensuring analyticity.  
 Besides the known resonance contributions, the background contributions are included via additional terms with poles below the $\pi$N threshold. 
 Due to the clear analytic and separable structure of the model, finding the pole positions in CMB model is trimmed down to the
  generalization of the dispersion integral for the channel propagator from real axes to the full complex energy plane, and this
   is a very well defined procedure. In practice, we instead use a very stable, and numerically much faster analytic continuation 
   method based on the Pietarienen expansion \cite{Pietarinen} in order to extrapolate the real valued channel propagator into the 
   complex energy plane. 

\subsection{Formulae}
Our current partial-wave analysis~\cite{Bat98} is based on the manifestly unitary, multichannel CMB approach of ref.~\cite{Cut79}. 
The most prominent property of this approach is  
analyticity of partial waves with respect to Mandelstam $s$ variable. In every discussion of partial-wave poles, analyticity plays a
 crucial role since the  
 poles are situated in a complex plane, away from physical region, and our measuring abilities are restricted to the real energy axis only.
  To gain any knowledge about the nature of partial-wave singularities would be impossible if partial waves were not analytic.
   Therefore, the ability to calculate pole positions is not just a benefit of the CMB model's  
analyticity but also a necessity for resonance extraction. In this approach, the resonance itself is considered to exist 
if there is an  associated partial-wave pole in the ``unphysical'' sheet. 
\\ \\ \noindent
We use the  multichannel $T$ matrix related to the scattering matrix $S$ as:
\begin{equation}
S_{ab}(s)= \delta_{ab} + 2\,i\,T_{ab}(s), 
\end{equation}
 where $\delta_{ab}$ is Kronecker delta symbol.  
The T-matrix matrix element is in the CMB model given as: 
\begin{eqnarray}
\label{eq:Tmatrix}
 T_{ab}^{JL}(s) 
= \sum_{i,j=1}^{N^{JL}}  f_a^{JL}(s)  \sqrt{\rho_a (s)}  \gamma_{ai}^{JL}G_{ij}^{JL}(s)\gamma_{jb}^{JL}\sqrt{\rho_b(s)}f_b^{JL}(s) & & \nonumber
\end{eqnarray}
where $a(b)$ represents the outgoing (incoming) channel. In our analyses we use $a,b=\pi N,\eta N,\pi^2 N$. The initial and final channel $b(a)$ 
couple through intermediate particles labeled $i$ and $j$. The factors $\gamma_{ia}$  are energy-independent parameters 
occurring graphically at the vertex between channel $a$ and intermediate particle $i$ and are determined by fitting procedure. Also occurring at
 each initial or final vertex is form factor $f_a^{JL}(s)$
\begin{equation}
\label{eq:ch-resvfun}
 f_a^{JL}(s)=\left(\frac{q_a}{Q_{1a}+\sqrt{Q_{2a}^{2}+q_a^2}}\right)^L,
\end{equation}
and phase-space factor $\rho_a(s)$
\begin{equation}
 \rho_a(s)=\frac{q_a(s)}{\sqrt{s}},
\end{equation}
where $s=W^2$ is a Mandelstam variable, and  $q_a(s)$ is the meson momentum for any of the three channels given as 
\begin{eqnarray}
 q_a(s) =  \frac{\sqrt{(s-(m+m_a)^2)(s-(m-m_a)^2)}}{2\sqrt{s}}. 
\end{eqnarray}

Furthermore, $L$ is the angular momentum in channel $a$, and $Q_{1a}$, $Q_{2a }$ are constants. The factor $f_a^{JL}(s)$ provides appropriate threshold behavior 
on the right-hand cut, and also produces a left-hand branch cut in the $s$ plane. Parameters $Q_{1a}$ and $Q_{2a}$ are chosen to determine the
 branch point and strength of the left-hand branch cut. In our analyses they have been taken to be the same, and are fixed to the mass of
 the channel meson $a$.

$G_{ij}^{JL}$ is a dressed propagator for partial wave $JL$ and particles $i$ and $j$, and may be written in terms of a diagonal bare propagator 
$G_{ij}^{0JL}$ and a self-energy matrix $\Sigma_{kl}^{JL}$ using Dyson equation
\begin{eqnarray}\label{Dyson}
 G_{ij}^{JL}(s) & = & G_{ij}^{0JL}(s)+\sum_{k,l=1}^{N^{JL}}G_{ik}^{0JL}(s)\Sigma_{kl}^{JL}(s)G_{ij}^{JL}(s).   
\end{eqnarray}
\\

The bare propagator 
\begin{equation}\label{bare}
 G_{ij}^{0JL}(s)=\frac{e_i\delta_{ij}}{s_i-s}
\end{equation}
has a pole at the real value $s_i$. The sign $e_i=\pm 1$ must be chosen to be positive for poles above the elastic threshold which correspond 
to resonance. 

The nonresonant background is described by a meromorphic function, in most of the cases consisting of two terms of the form 
(\ref{bare}) with pole positions below $\pi N$ threshold. For that case, the signs of the terms are opposite. The positive sign correspond to the
 repulsive and the negative sign to the attractive potential. In principle the number of poles can be increased arbitrarily 
 (see the next subsection on background representation), but in reality the number is never larger than three. 

$\Sigma_{kl}^{JL}$  is the self-energy term for the particle propagator
\begin{equation}
 \Sigma_{kl}^{JL}(s)=\sum_a \gamma_{ka}^{JL} \cdot \Phi_a^{JL}(s) \cdot \gamma_{la}^{JL}
\end{equation}
The $\Phi_{a}^{JL}(s)$ are called ``channel propagators''. They are constructed in an approximation that threats each channel as containing
only two particles. We require that $T_{ab}^{JL}$ have, in all channels, correct unitarity and analyticity properties consistent with a 
 quasi-two-body approximation. 
 
 The imaginary part of $\Phi_a^{JL}(s)$ is the effective phase-space factor for channel $a$:
\begin{equation}
\label{eq:F}
 \mathrm{Im} \ \Phi_a^{JL}(s)=[f_a^{JL}(s)]^2\rho_a(s).
\end{equation}

\begin{figure*}[!tb]\begin{center}
\includegraphics[width=8.cm]{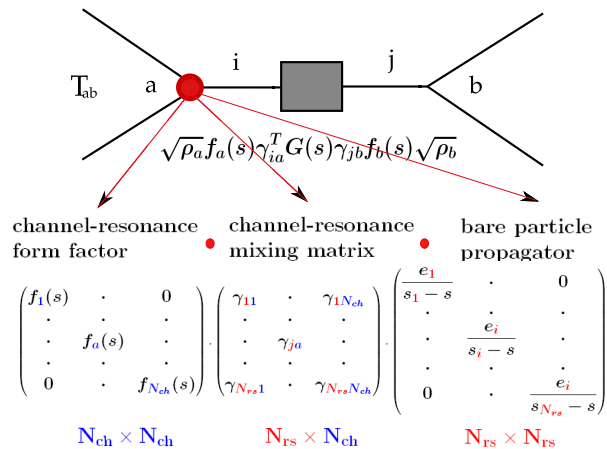}\\
\end{center} 
 \caption{Parameterization of channel-intermediate particle vertex in CMB model.}
 \label{Figure1}
\end{figure*}

The channel propagator is evaluated on the real axes only
\begin{equation}\label{eq:imphi}
\mathrm{Im}\, \Phi(x)=\frac{\left[q(x)\right]^{2L+1}}{\sqrt{x}\,\left\{Q_1+\sqrt{Q_2^2+\left[q(x)\right]^2}\right\}^{2L}}, \vspace*{0.1cm}
\end{equation}
where by $x$ we stress the fact that values are on the real axes. 
The real part of $\Phi_a^{JL}(x)$ is calculated using a subtracted dispersion relation

\begin{equation}\label{eq:DR}
\mathrm{Re} \, \Phi(x)=\frac{x-x_a}{\pi}\,\,\mathrm{P}\int_{x_a}^{\infty}\frac{\mathrm{Im}\,\Phi(x')\,dx'}{(x'-x)(x'-x_a)}.
\end{equation}
where $x_a=(m+m_a)^2$. 
For better understanding, the structure of the channel-intermediate particle form factor is given in Fig. \ref{Figure1}.\\
 
We give a matrix form of the final T matrix as defined in Eq.~(\ref{eq:Tmatrix}):
\begin{widetext}
 \begin{eqnarray}
      \hat{T}(s) = 
	  \sqrt{Im  \hat{\Phi}(s)} \cdot
    \hat{\gamma}^{\rm T} \cdot  \frac{\hat{G}_{0}(s)}{I- \left[\hat{\gamma} \cdot \hat{\Phi}(s) \cdot \hat{\gamma}{\rm ^T}\right] \cdot
	 \hat{G}_{0}(s)} \cdot  \hat{\gamma}  \cdot \sqrt{Im  \hat{\Phi} (s) } 
\label{eq:final}
\end{eqnarray}
\end{widetext}
\subsection{General idea}
In this paper we propose to use a method based on coupled-channel formalism, apply it to all partial wave data and partial wave amplitudes available ``on the market", and  
simultaneously analyze the underlying analytic structure. We have decided to use \emph{only one model} to extract pole positions from \emph{all published partial waves 
analyses} in order to evade the model assumptions of each approach, and compare the results on the same footing. And we have chosen the T-matrix Carnagie-Melon-Berkeley (CMB) 
model. In other words, we take all sets of partial-wave amplitudes, accept them as nothing else but good representations of all analyzed experimental data, and extract the 
poles which are required by the CMB method. In this manner, all errors due to different analytic continuations of different models are avoided, and the only remaining error is 
the precision of CMB method itself (see ref. \cite{Hedim2010}). Of course, we shall compare the obtained poles with the poles of each individual publication, and draw certain 
conclusions about features of individual methods as well.

\subsection{Data base}

We start with a collection of data in which one part is fully available in the literature \cite{KH80,Arn04,GWUWEB}, and numeric values for the second part are provided  by the  
authors (private communication refs. \cite{Giessen,Diaz07,Dur08,Che07,Juelich}). \\

We have analyzed the following  PWA amplitudes:
\begin{enumerate}
\item Karlsruhe-Helsinki (KH80) \cite{KH80} - $\pi$N elastic; \\
As the influence of inelastic channels is in KH80 formalism introduced through forward and fixed \textit{cms} scattering  dispersion relations, KH80 does not offer any 
inelastic channel amplitudes to be fitted.  However, as we know that inelastic channels are extremely important in CMB formalism to ensure the stability of solutions (see 
following chapter and ref. \cite{Ceci06}), we have decided to constrain elastic KH80 amplitudes with $\pi$N $\rightarrow \eta$N WI08 amplitudes which fairly correctly depict 
the world agreement  of the $\eta$N channel amplitudes at lower energies - see Fig. \ref{Fig_podaci_inelastic}.
\item VPI/GWU - $\pi$N elastic and $\pi$N $\rightarrow \eta$N; \\
We have used single energy solutions (GWU-SES) \cite{GWUWEB}, and energy dependent solutions (WI08) \cite{Arn04,GWUWEB}.
\item Giessen  \cite{Giessen} - $\pi$N elastic and $\pi$N $\rightarrow \eta$N
\item EBAC - We have used two sets of PW solutions. Single-channel fit solution ($\pi$N elastic fitted) --- EBAC07 \cite{Diaz07}, and two channel fit solution ($\pi$N elastic 
and $\pi$N $\rightarrow \eta$N fitted) --- EBAC08 \cite{Dur08} with the \mbox{$\pi$N $\rightarrow \eta$N}  normalization adjusted in accordance with M. D\"{o}ring and B. Diaz 
\cite{Duering2009}.
with M. D\"{o}ring and B. Diaz \cite{Duering2009}
\item J\"{u}lich   \cite{Juelich} - $\pi$N elastic and $\pi$N $\rightarrow \eta$N 
\item Dubna-Mainz-Taipei (DMT) \cite{Che03,Che07} - $\pi$N elastic and   $\pi$N $\rightarrow \eta$N 
\end{enumerate}

\subsection{Fitting procedure}
We have used three channel CMB formalism with $\pi$N and $\eta$N physical channels, and the third, effective two body channel to account for unitarity. 
We start with a minimal number of bare poles, and increase their number as long as the quality of the fit, measured by the lowest reduced $\chi^2$ value, could not be improved. 
In addition, a visual resemblance of the fitting curve to the data set in totality was used as a rule of thumb; i.e., we rejected those solutions that had a tendency  to 
accommodate for the rapidly varying data points, regardless of the $\chi^2$ value. When the optimal number of poles is reached, we claim that we have found all partial wave 
pole solutions given by the chosen data set. As our criteria (minimal reduced $\chi^2$ value and visual resemblance) are not extremely rigid, we have to differentiate between 
the two categories of poles: those which are seen with almost complete certainty, and those which are only consistent with the chosen set of data. The poles whose addition 
significantly improve the reduced $\chi^2$ value fall into the first category, those which improve the reduced $\chi^2$ value only marginally fall into the second one. It is 
interesting to note that in the latter case a number of almost equivalent, indistinguishable solutions for the questionable pole may be found. 

\section{Results and Discussion}

 The intention of this article is to use only one method, Zagreb realization of CMB model, to extract pole positions from a world collection of partial wave data and partial 
wave amplitudes. As a test case, we  do it for the S$_{11}$ partial wave only. We use three channel model, with two measured channels  $\pi$N, $\eta$N, and the third channel 
$\pi^2$N,  which effectively represents all other inelastic channels, and ``takes care of '' unitarity. \\

We extract pole positions from all available PWA and make a comprehensive analyses. We analyze the number of poles needed for a given partial wave, we discuss the importance of 
inelastic channels. 

\subsection{Methodology}
The main feature of CMB multi-resonance, multi-channel model is a good control over determining the number of bare poles, and deducing the importance of number of fitted 
channels. 
\\ \\ \noindent 
\underline{\textit{Importance of inelastic channels}} \\

The elastic $\pi$N scattering channel is the best measured and the most confident channel, so in all cases it is the pillar of the obtained partial wave amplitudes. Most of the 
information about the energy dependent structure of all solutions is coming from this channel, and it is expected that corrections are coming from other channels. Therefore, it 
is carrying the heaviest weight for obtaining final results.  \\

At  this point we are bound to address one specific point in more details. \\

In ref. \cite{Ceci06} we have discussed the continuum ambiguity problem in coupled-channel formalisms. Namely, once the inelastic channels are opened, it turns out that the 
differential cross sections themselves are not sufficient to determine the scattering amplitude. Let us illustrate why.  If differential cross section $d\sigma / d\Omega$ is 
given by $|F|^2$, then the new function \emph{\~{F}}~$= e^{i\Phi} F$ gives exactly the same cross section. It should be remarked that this phase uncertainty has nothing to do 
with the non-observable phase of wave functions in quantum mechanics. The asymptotic wave functions at large distances from the scattering center may be written as $\Psi(x) 
\approx e^{i \cdot k \cdot x} + F(\theta) \frac{e^{i \cdot r \cdot }}{r}$, $ r \rightarrow \infty$, so the phase of scattering amplitude is the \emph{relative} phase of the 
incident and scattered wave. This phase has observable consequences in situations where multiple scattering occurs, and the continuum ambiguity is created. In the elastic 
region, the unitarity \textit{relates} real and imaginary parts of each partial wave, and the consequence is a constraint which effectively removes this "continuum" ambiguity, 
and leaves potentially only a discreet one. The partial wave must lie \emph{on} the unitary circle. However, as soon as the inelastic threshold opens, unitarity provides 
\emph{only} an inequality: $ |1+2 \ i \ F_l|^2 \leq 1 \Longrightarrow {\rm Im F}_l = |F_l|^2+ I_l$, where $I_l = \frac{1}{4}(1 - e^{- \, 4 \, {\rm Im} \, \delta_l})$. 
Therefore, each partial wave must lie \emph{upon or inside} its unitary circle, and not on it. A whole family of functions $\Phi$, of limited magnitude but of infinite variety 
of functional forms which satisfy the required conditions, does exist. However, in spite that they contain a continuum infinity of points, they are limited in extent. Thus, the 
\textit{islands of ambiguity} are created. \\

In ref. \cite{Ceci06} we have shown that including inelastic channels into the analysis is a natural way for eliminating continuum ambiguities. 
We have concluded that, by fitting only elastic channel, some of the resonant states which dominantly couple to inelastic channels might remain unrevealed, and we had to fit as 
many channels as possible. In the present paper we apply the following strategy: we shall first fit elastic channel only, and show the poles we reveal. Then, we shall repeat 
the fit by fitting two channel processes, $\pi$N elastic and $\pi$N $\rightarrow \eta$N data when available, and see how the number of poles, and their quantitative values 
change.  \\

The problem we are facing is the low quality input for the $\eta$N channel, because $\pi$N $\rightarrow \eta$N partial waves are in principle not well known. Anyway, as 
\textbf{\textit{a final result}}, we have to accept the solution for which both channels are \textbf{\textit{reasonably well}} fitted despite the low quality of the $\eta$N 
channel data.  
\\ \\ \\ \noindent 
\underline{\textit{Determining the optimal number of poles}} \\

In CMB formalism the number of poles is a starting parameter. That in practice means that when fitting, we start with a  minimal set of poles: one resonant and two for the 
background. Then we increase the number of resonant poles until the satisfactory fit is achieved, i.e., until the quality of the fit, measured by the  reduced $\chi^2$ value, 
could not be  improved. In addition, a visual resemblance of the fitting curve to the data set as a whole is used as a rule of thumb: we reject  all those solutions which have 
a tendency  to accommodate for the rapidly  varying data points regardless of the $\chi^2$ value. 

In such a way we estimate the number of bare poles needed by our model, what in most  cases corresponds to the number of resonant states. Observe that this is not so for 
dynamic resonances, i.e., for the dressed resonant states which do not have a corresponding bare pole. Therefore, what we compare \textbf{\textit{is not}} the number of bare 
poles, but \textbf{\textit{the number of dressed ones}}. (For a more extensive discussion on dynamic resonant states in Zagreb CMB model see ref. \cite{Ceci08}.)

\subsection{Fits}
We first  fit $\pi$N elastic channel only. In accordance with the afore considerations, we first want to determine which resonances are we well determined only by this channel, 
and later on we want to see how much the inclusion of $\eta$N channel will modify the obtained result. 
\subsubsection{$\pi$N  elastic channel only} 
\begin{figure}[!h]
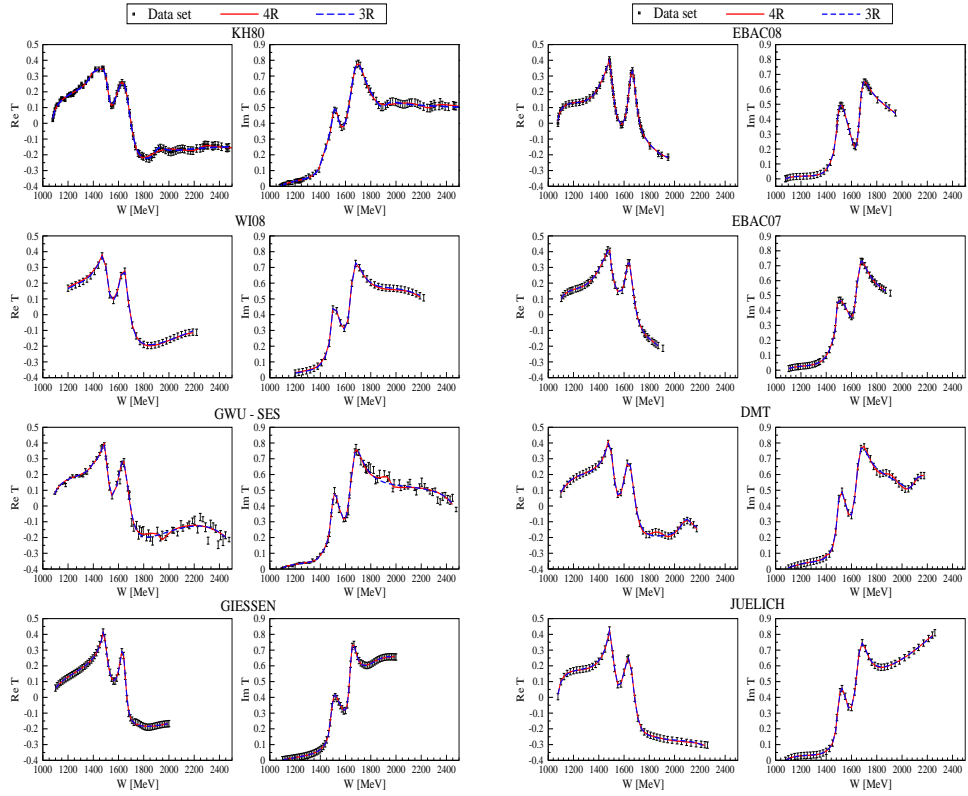

\begin{center}
\subfigure{\includegraphics[width=6cm,height=10.5cm]{Single_Fit1}} \hspace*{0.5cm}
\subfigure{\includegraphics[width=6cm,height=10.5cm]{Single_Fit2}}
\end{center} 
 \caption{Agreement of 3R and 4R CMB curves with ``world input" for single-channel fit.}
 \label{Fig_podaci_elastic}
\end{figure}

We show the result of the fit in Table \ref{Table1}. The quality of the fit is shown in Fig. \ref{Fig_podaci_elastic}.

\begin{table*}[!H]
\caption{\footnotesize ``World collection" of poles for the single-channel fit, three and four resonant case. }
\begin{center}
\begin{tabular}{|c|c|c|cccc|cccc|c|}
\hline \hline 
&\multirow{3}{*}{Fitted} & \multirow{2}{*}{Number} &  \multicolumn{4}{c|}{ Bare poles} & \multicolumn{4}{c|}{ Dressed poles } & \\ [0ex] \cline{4-11}
 ANALYSES \ \   & \multirow{3}{*}{channel} & \multirow{2}{*}{of} &$ \mathrm{W_{s_1}}$ & $\mathrm{W_{s_2}}$ &  $\mathrm{W_{s_3}}$   & $\mathrm{W_{s_4}}$ & {\footnotesize { 
$\begin{pmatrix}
\mathrm{Re W} \\ \mathrm{-2Im W} 
\end{pmatrix}$}} & {\footnotesize { $\begin{pmatrix}
\mathrm{Re W}\\ \mathrm{-2Im W} 
\end{pmatrix}$}} & {\footnotesize { $\begin{pmatrix}
\mathrm{Re W}\\ \mathrm{-2Im W} 
\end{pmatrix}$}}  & {\footnotesize { $\begin{pmatrix}
\mathrm{Re W}\\ \mathrm{-2Im W} 
\end{pmatrix}$}}  & $\chi_R^2$  \\ [0ex]
 &  & \multirow{1}{*}{resonances} & & \multicolumn{2}{c}{$\mathrm{MeV}$}  &  & & \multicolumn{2}{c}{$\mathrm{MeV}$}  & &   \\   \hline \hline
 \multirow{3}{*}{KH80} &\multirow{3}{*}{$\pi N\rightarrow \pi N$}&\footnotesize{3}& \footnotesize{1516} & \footnotesize{1638} & \footnotesize{1880} &- &{\footnotesize { 
$\begin{pmatrix}
1513\\ 71
\end{pmatrix}$}} & {\footnotesize { $\begin{pmatrix}
1661\\ 148
\end{pmatrix}$}} &	{\footnotesize { $\begin{pmatrix}
1903\\ 90
\end{pmatrix}$}} &-& \footnotesize{0.209}\\
 &&\footnotesize{4}& \footnotesize{1488} & \footnotesize{1656} & \footnotesize{1713} & \footnotesize{2266}&{\footnotesize { $\begin{pmatrix}
1513\\ 113
\end{pmatrix}$}} & {\footnotesize { $\begin{pmatrix}
1670\\ 194
\end{pmatrix}$}} &	{\footnotesize { $\begin{pmatrix}
1833\\ 703
\end{pmatrix}$}}&	{\footnotesize { $\begin{pmatrix}
2263\\ 138
\end{pmatrix}$}} &  \footnotesize{0.206}  \\[0ex]\hline
 \multirow{3}{*}{WI08}&\multirow{3}{*}{$\pi N\rightarrow \pi N$}&\footnotesize{3}& \footnotesize{1481} & \footnotesize{1657} & \footnotesize{3767} &-& {\footnotesize { 
$\begin{pmatrix}
1492\\ 89
\end{pmatrix}$}} & {\footnotesize { $\begin{pmatrix}
1646\\ 95
\end{pmatrix}$}} &	{\footnotesize { $\begin{pmatrix}
2684\\ 822
\end{pmatrix}$}} &	-&  \footnotesize{0.043} \\[0ex] 
&& \footnotesize{4}&\footnotesize{1513}& \footnotesize{1624}& \footnotesize{1686} & \footnotesize{2517}&{\footnotesize { $\begin{pmatrix}
1495\\ 105
\end{pmatrix}$}} & {\footnotesize { $\begin{pmatrix}
1647\\ 81 
\end{pmatrix}$}} &	{\footnotesize { $\begin{pmatrix}
1658\\ 255
\end{pmatrix}$}}& {\footnotesize { $\begin{pmatrix}
2396\\ 139
\end{pmatrix}$}}&  \footnotesize{0.012}  \\ [0ex] \hline

\multirow{3}{*}{GWU-SES}&\multirow{3}{*}{$\pi N\rightarrow \pi N$}& \footnotesize{3}&\footnotesize{1514} & \footnotesize{1645} & \footnotesize{2919} & - &{\footnotesize { 
$\begin{pmatrix}
1500\\ 106
\end{pmatrix}$}} & {\footnotesize { $\begin{pmatrix}
1646\\ 119
\end{pmatrix}$}} &	\centering {\footnotesize { $\begin{pmatrix}
2598\\ 210
\end{pmatrix}$}} &- & \footnotesize{2.252} \\[0ex]
&&\footnotesize{4}& \footnotesize{1517} & \footnotesize{1650} & \footnotesize{1928} &\footnotesize{3768}&{\footnotesize { $\begin{pmatrix}
1505\\ 97
\end{pmatrix}$}} & {\footnotesize { $\begin{pmatrix}
1651\\ 119
\end{pmatrix}$}} &	{\footnotesize { $\begin{pmatrix}
1944\\ 74
\end{pmatrix}$}} &	{\footnotesize { $\begin{pmatrix}
2633\\ 345
\end{pmatrix}$}}&  \footnotesize{2.116} \\[0ex]\hline
\multirow{3}{*}{GIESSEN}&\multirow{3}{*}{$\pi N\rightarrow \pi N$}& \footnotesize{3}& \footnotesize{1464} & \footnotesize{1616} & \footnotesize{1731} & - &{\footnotesize { 
$\begin{pmatrix}
1484\\ 82
\end{pmatrix}$}} & {\footnotesize { $\begin{pmatrix}
1641\\ 65
\end{pmatrix}$}} &	\centering {\footnotesize { $\begin{pmatrix}
1861\\ 811
\end{pmatrix}$}} &- & \footnotesize{0.062} \\[0ex]
&&\footnotesize{4}& \footnotesize{1474} & \footnotesize{1635} & \footnotesize{1718}&\footnotesize{2674}&{\footnotesize { $\begin{pmatrix}
1482\\ 82
\end{pmatrix}$}} & {\footnotesize { $\begin{pmatrix}
1642\\ 65
\end{pmatrix}$}} &	{\footnotesize { $\begin{pmatrix}
1851\\ 456
\end{pmatrix}$}} &	{\footnotesize { $\begin{pmatrix}
2249\\ 287
\end{pmatrix}$}}&  \footnotesize{0.061} \\[0ex]\hline 
\multirow{3}{*}{JUELICH}&\multirow{3}{*}{$\pi N\rightarrow \pi N$}& \footnotesize{3}&\footnotesize{1518} & \footnotesize{1656} & \footnotesize{2177} & - &{\footnotesize { 
$\begin{pmatrix}
1528\\ 95
\end{pmatrix}$}} & {\footnotesize { $\begin{pmatrix}
1653\\ 110
\end{pmatrix}$}} &	\centering {\footnotesize { $\begin{pmatrix}
2335\\ 372
\end{pmatrix}$}} & -& \footnotesize{0.046} \\[0ex]
&&\footnotesize{4}& \footnotesize{1511} & \footnotesize{1636} & \footnotesize{1719} &\footnotesize{2241}&{\footnotesize { $\begin{pmatrix}
1516\\ 121
\end{pmatrix}$}} & {\footnotesize { $\begin{pmatrix}
1654\\ 118
\end{pmatrix}$}} &	{\footnotesize { $\begin{pmatrix}
1665\\ 411
\end{pmatrix}$}} &	{\footnotesize { $\begin{pmatrix}
2335\\ 403
\end{pmatrix}$}}&  \footnotesize{0.018}\\[0ex]\hline 
\multirow{3}{*}{EBAC07}&\multirow{3}{*}{$\pi N\rightarrow \pi N$}& \footnotesize{3}&\footnotesize{1466} & \footnotesize{1641} & \footnotesize{2518} &  -&{\footnotesize { 
$\begin{pmatrix}
1498\\ 123
\end{pmatrix}$}} & {\footnotesize { $\begin{pmatrix}
1641\\ 89
\end{pmatrix}$}} &	\centering {\footnotesize { $\begin{pmatrix}
2215\\ 767
\end{pmatrix}$}} & -& \footnotesize{0.028} \\[0ex]
&&\footnotesize{4}& \footnotesize{1483}& \footnotesize{1643} & \footnotesize{1702} &\footnotesize{2237}&{\footnotesize { $\begin{pmatrix}
1502\\ 139
\end{pmatrix}$}} & {\footnotesize { $\begin{pmatrix}
1638\\ 81
\end{pmatrix}$}} &	{\footnotesize { $\begin{pmatrix}
1700\\ 408
\end{pmatrix}$}} &	{\footnotesize { $\begin{pmatrix}
1862\\ 691
\end{pmatrix}$}}&  \footnotesize{0.012} \\[0ex]\hline 
\multirow{3}{*}{EBAC08}&\multirow{3}{*}{$\pi N\rightarrow \pi N$}& \footnotesize{3}&\footnotesize{1515} & \footnotesize{1673} & \footnotesize{1826} &  -&{\footnotesize { 
$\begin{pmatrix}
1483\\ 123
\end{pmatrix}$}} & {\footnotesize { $\begin{pmatrix}
1662\\ 80
\end{pmatrix}$}} &	\centering {\footnotesize { $\begin{pmatrix}
1873\\ 219
\end{pmatrix}$}} &- & \footnotesize{0.029} \\[0ex]
&&\footnotesize{4}& \footnotesize{1512} & \footnotesize{1667} & \footnotesize{1980} &\footnotesize{3784}&{\footnotesize { $\begin{pmatrix}
1492\\ 114
\end{pmatrix}$}} & {\footnotesize { $\begin{pmatrix}
1661\\ 81
\end{pmatrix}$}} &	{\footnotesize { $\begin{pmatrix}
1804\\ 1113
\end{pmatrix}$}} &	{\footnotesize { $\begin{pmatrix}
2189\\ 637
\end{pmatrix}$}}&  \footnotesize{0.027} \\[0ex]\hline 

\multirow{3}{*}{DMT}&\multirow{3}{*}{$\pi N\rightarrow \pi N$}& \footnotesize{3}&\footnotesize{1495} & \footnotesize{1643} & \footnotesize{2047} &-  &{\footnotesize { 
$\begin{pmatrix}
1486\\ 81
\end{pmatrix}$}} & {\footnotesize { $\begin{pmatrix}
1640\\ 103
\end{pmatrix}$}} &	\centering {\footnotesize { $\begin{pmatrix}
2080\\ 100
\end{pmatrix}$}} & -& \footnotesize{0.246} \\[0ex]
&&\footnotesize{4}& \footnotesize{1507} & \footnotesize{1647} & \footnotesize{1850} &\footnotesize{2100}&{\footnotesize { $\begin{pmatrix}
1508\\ 139
\end{pmatrix}$}} & {\footnotesize { $\begin{pmatrix}
1643\\ 134
\end{pmatrix}$}} &	{\footnotesize { $\begin{pmatrix}
1892\\ 203
\end{pmatrix}$}} &	{\footnotesize { $\begin{pmatrix}
2100\\ 212
\end{pmatrix}$}}&  \footnotesize{0.083} \\[0ex]\hline \hline
\multirow{3}{*}{Averages} &    &  \footnotesize{3} &  &       &      &    &{\footnotesize { $\begin{pmatrix}
1498(16) \\ 96(20)
\end{pmatrix}$}} & {\footnotesize { $\begin{pmatrix}
1649(9) \\ 101(25)
\end{pmatrix}$}} &	\centering {\footnotesize { $\begin{pmatrix}
2194(324) \\ 424(324)
\end{pmatrix}$}} & -&  \\[0ex]
  &  & \footnotesize{4} &  &  &  &  &{\footnotesize { $\begin{pmatrix}
1502(12) \\ 114(20)
\end{pmatrix}$}} & {\footnotesize { $\begin{pmatrix}
1651(11) \\ 109(42)
\end{pmatrix}$}} &	{\footnotesize { $\begin{pmatrix}
1793(108)\\ 453(325)
\end{pmatrix}$}} &	{\footnotesize { $\begin{pmatrix}
2253(224) \\ 356(212)
\end{pmatrix}$}}&   \\[0ex]\hline  \hline
\end{tabular}
\end{center}
\label{Table1}
\end{table*}

\clearpage
\subsubsection{$\pi$N  elastic and $\pi$N $\rightarrow \eta$N data} 

As we have already mentioned,  $\pi$N $\rightarrow \eta$N data are rather old and vague, so the corresponding partial waves are poorly determined. Anyway, each analyzed PWA 
solution of our ``world collection", with the exception of KH80 does offer some results for that channel, and  we have consistently used it in the two channels fit. The only 
exception, KH80 amplitudes, do not have a corresponding $\eta$N channel. We have been tempted to omit KH80 amplitudes from the coupled-channel analysis, but due to its 
extremely good analytical constraints, we have decided to keep it in some form. Instead of KH80 $\eta$N  channel, we have used the WI08 VPI/GWU solution believing that the 
S$_{11}$ $\eta$N channel amplitudes are confidently well known in the energy range  \mbox{s $\leq$ 3 GeV$^2$}  (\mbox{$T_{lab} \leq 800 $ MeV}), and in that range the WI08 
VPI/GWU solution is a good numeric representation of a ``world collection average".

\begin{figure}
\begin{center}
\includegraphics[width=14.cm]{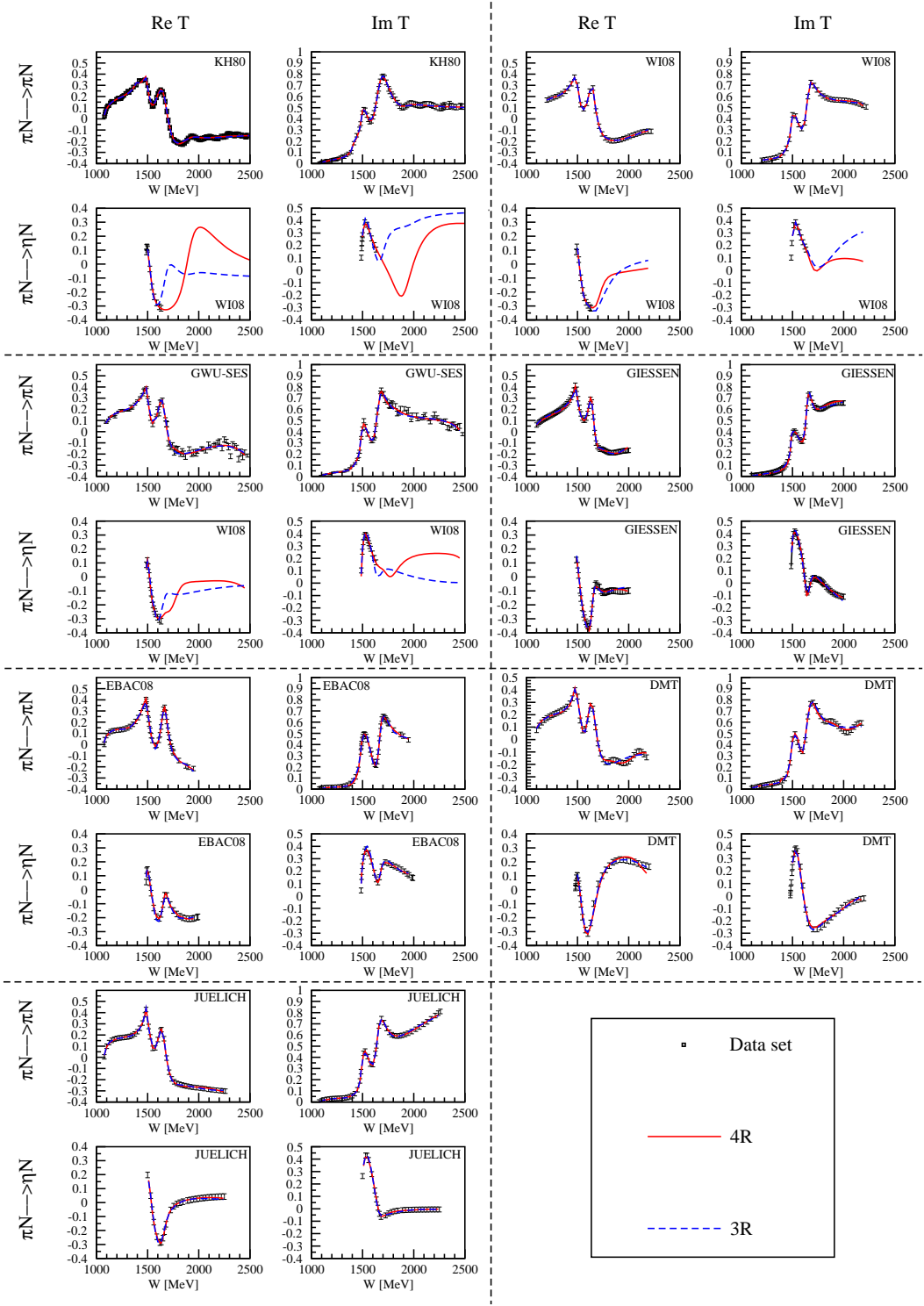}\\
\end{center} 
 \caption{Agreement of 3R and 4R CMB curves with ``world input" for two-channel fit.}
 \label{Fig_podaci_inelastic}
\end{figure}

We show the result of the fit in Table~\ref{Table2}. The quality of the fit is shown in Fig. \ref{Fig_podaci_inelastic}.  

\begin{table*}[!h]
\caption{\footnotesize ``World collection" of poles for the two channels fit, three and four resonant case. }
\begin{center}
\begin{tabular}{|cc|c|cccc|cccc|c|}
\hline \hline 
\multicolumn{2}{|c|}{\multirow{2}{*}{ANALYSIS}}&Number&  \multicolumn{4}{c|}{ Bare poles} & \multicolumn{4}{c|}{ Dressed poles } & \\ \cline{4-12}
\multicolumn{2}{|c|}{\multirow{3}{*}{(Fitted channels)}}&  of&$\mathrm{W_{s_1}}$ & $\mathrm{W_{s_2}}$ &  $\mathrm{W_{s_3}}$   & $\mathrm{W_{s_4}}$ & {\footnotesize { 
$\begin{pmatrix}
\mathrm{Re W}\\ \mathrm{-2Im W} 
\end{pmatrix}$}} & {\footnotesize { $\begin{pmatrix}
\mathrm{Re W}\\ \mathrm{-2Im W} 
\end{pmatrix}$}} & {\footnotesize { $\begin{pmatrix}
\mathrm{Re W}\\ \mathrm{-2Im W} 
\end{pmatrix}$}}  & {\footnotesize { $\begin{pmatrix}
\mathrm{Re W}\\ \mathrm{-2Im W} 
\end{pmatrix}$}}  & $\chi_R^2$  \\ 
\multicolumn{2}{|c|}{}&resonances && \multicolumn{2}{c}{$\mathrm{MeV}$}  &  & & \multicolumn{2}{c}{$\mathrm{MeV}$}  & &  \\ \hline \hline
 KH80 & WI08 & \footnotesize {3} & \footnotesize{1517} & \footnotesize{1637} & \footnotesize{1865} &- &{\footnotesize { $\begin{pmatrix}
1511\\ 113
\end{pmatrix}$}} & {\footnotesize { $\begin{pmatrix}
1670\\ 163
\end{pmatrix}$}} &	{\footnotesize { $\begin{pmatrix}
1923\\ 328
\end{pmatrix}$}} &-& \footnotesize{0.391}\\[0ex]
\scriptsize {($\pi N\rightarrow \pi N$)}&\scriptsize{ ($\pi N \rightarrow \eta N$)}&\footnotesize{4}& \footnotesize{1504} & \footnotesize{1610} & \footnotesize{1751} 
&\footnotesize{2045}&{\footnotesize { $\begin{pmatrix}
1492\\ 122
\end{pmatrix}$}} & {\footnotesize { $\begin{pmatrix}
1650\\ 163
\end{pmatrix}$}} &	{\footnotesize { $\begin{pmatrix}
1892\\ 235
\end{pmatrix}$}}&	{\footnotesize { $\begin{pmatrix}
1951\\ 555
\end{pmatrix}$}} &  \footnotesize{0.307}  \\[0ex]\hline
\multicolumn{2}{|c|}{WI08}&\footnotesize{3}& \footnotesize{1514} & \footnotesize{1626} & \footnotesize{1722} &- &{\footnotesize { $\begin{pmatrix}
1499\\ 114
\end{pmatrix}$}} & {\footnotesize { $\begin{pmatrix}
1652\\ 102
\end{pmatrix}$}} &	{\footnotesize { $\begin{pmatrix}
1718\\ 449
\end{pmatrix}$}} &-& \footnotesize{0.127}\\[0ex]
\scriptsize {($\pi N\rightarrow \pi N$)}&\scriptsize{ ($\pi N \rightarrow \eta N$)} &\footnotesize{4}& \footnotesize{1513} & \footnotesize{1630} & \footnotesize{1701} 
&\footnotesize{2611}&{\footnotesize { $\begin{pmatrix}
1495\\ 113
\end{pmatrix}$}} & {\footnotesize { $\begin{pmatrix}
1651\\ 87
\end{pmatrix}$}} &	{\footnotesize { $\begin{pmatrix}
1697\\ 204
\end{pmatrix}$}}&	{\footnotesize { $\begin{pmatrix}
2422\\ 241
\end{pmatrix}$}} &  \footnotesize{0.031} \\[0ex]\hline
GWU-SES&WI08&\footnotesize{3}& \footnotesize{1519} & \footnotesize{1662} & \footnotesize{3190} &- &{\footnotesize { $\begin{pmatrix}
1503\\ 172
\end{pmatrix}$}} & {\footnotesize { $\begin{pmatrix}
1642\\ 127
\end{pmatrix}$}} &	{\footnotesize { $\begin{pmatrix}
2618\\ 270
\end{pmatrix}$}} &-& \footnotesize{2.451}\\[0ex]
\scriptsize{($\pi N\rightarrow \pi N$)}&\scriptsize{ ($\pi N \rightarrow \eta N$)}&\footnotesize{4}& \footnotesize{1512}& 1643 & \footnotesize{1743} 
&\footnotesize{2827}&{\footnotesize { $\begin{pmatrix}
1503\\ 97
\end{pmatrix}$}} & {\footnotesize { $\begin{pmatrix}
1659\\ 111
\end{pmatrix}$}} &	{\footnotesize { $\begin{pmatrix}
1756\\ 210
\end{pmatrix}$}}&	{\footnotesize { $\begin{pmatrix}
2569\\ 173
\end{pmatrix}$}} &  \footnotesize{2.011}  \\[0ex]\hline
\multicolumn{2}{|c|}{GIESSEN}&\footnotesize{3}& \footnotesize{1515} & \footnotesize{1636} & \footnotesize{1720} &- &{\footnotesize { $\begin{pmatrix}
1472\\ 176
\end{pmatrix}$}} & {\footnotesize { $\begin{pmatrix}
1650\\ 81
\end{pmatrix}$}} &	{\footnotesize { $\begin{pmatrix}
1692\\191 
\end{pmatrix}$}} &-& \footnotesize{0.437}\\[0ex]
\scriptsize{($\pi N\rightarrow \pi N$)}&\scriptsize{ ($\pi N \rightarrow \eta N$)}&\footnotesize{4}& \footnotesize{1509}& \footnotesize{1632} & \footnotesize{1728} 
&\footnotesize{2202}&{\footnotesize { $\begin{pmatrix}
1471\\ 212
\end{pmatrix}$}} & {\footnotesize { $\begin{pmatrix}
1640\\ 73
\end{pmatrix}$}} &	{\footnotesize { $\begin{pmatrix}
1738\\ 263
\end{pmatrix}$}}&	{\footnotesize { $\begin{pmatrix}
2215\\ 246
\end{pmatrix}$}} &  \footnotesize{0.351}  \\[0ex]\hline
\multicolumn{2}{|c|}{JUELICH} &\footnotesize{3}& \footnotesize{1514} & \footnotesize{1601} & \footnotesize{1725} &- &{\footnotesize { $\begin{pmatrix}
1521\\ 212
\end{pmatrix}$}} & {\footnotesize { $\begin{pmatrix}
1649\\ 127
\end{pmatrix}$}} &	{\footnotesize { $\begin{pmatrix}
1643\\644 
\end{pmatrix}$}} &-& \footnotesize{0.198}\\[0ex]
\scriptsize{($\pi N\rightarrow \pi N$)}&\scriptsize{ ($\pi N \rightarrow \eta N$)}&\footnotesize{4}& \footnotesize{1513}& \footnotesize{1566} & \footnotesize{1663} 
&\footnotesize{2048}&{\footnotesize { $\begin{pmatrix}
1514\\ 142
\end{pmatrix}$}} & {\footnotesize { $\begin{pmatrix}
1633\\ 141
\end{pmatrix}$}} &	{\footnotesize { $\begin{pmatrix}
1645\\ 112
\end{pmatrix}$}}&	{\footnotesize { $\begin{pmatrix}
2197\\ 977
\end{pmatrix}$}} &  \footnotesize{0.074}  \\[0ex]\hline
\multicolumn{2}{|c|}{EBAC08}&\footnotesize{3}& \footnotesize{1518} & \footnotesize{1670} & \footnotesize{1883} &- &{\footnotesize { $\begin{pmatrix}
1526\\ 179
\end{pmatrix}$}} & {\footnotesize { $\begin{pmatrix}
1665\\ 126
\end{pmatrix}$}} &	{\footnotesize { $\begin{pmatrix}
1927\\ 347
\end{pmatrix}$}} &-& \footnotesize{0.651}\\[0ex]
 \scriptsize{($\pi N\rightarrow \pi N$)}&\scriptsize{ ($\pi N \rightarrow \eta N$)}&\footnotesize{4}& \footnotesize{1495} & \footnotesize{1618} & \footnotesize{1693} 
&\footnotesize{1888}&{\footnotesize { $\begin{pmatrix}
1493\\ 174
\end{pmatrix}$}} & {\footnotesize { $\begin{pmatrix}
1672\\ 87
\end{pmatrix}$}} &	{\footnotesize { $\begin{pmatrix}
1696\\ 122
\end{pmatrix}$}}&	{\footnotesize { $\begin{pmatrix}
1911\\ 107
\end{pmatrix}$}} &  \footnotesize{0.216}  \\[0ex]\hline
\multicolumn{2}{|c|}{DMT} &\footnotesize{3}& \footnotesize{1516} & \footnotesize{1657} & \footnotesize{2169} &- &{\footnotesize { $\begin{pmatrix}
1551\\ 160
\end{pmatrix}$}} & {\footnotesize { $\begin{pmatrix}
1638\\ 158
\end{pmatrix}$}} &	{\footnotesize { $\begin{pmatrix}
2378\\ 1070
\end{pmatrix}$}} &-& \footnotesize{1.186}\\[0ex]
 \scriptsize{($\pi N\rightarrow \pi N$)}&\scriptsize{ ($\pi N \rightarrow \eta N$)}&\footnotesize{4}& \footnotesize{1476} & \footnotesize{1606} & \footnotesize{1705} 
&\footnotesize{2104}&{\footnotesize { $\begin{pmatrix}
1546\\ 151
\end{pmatrix}$}} & {\footnotesize { $\begin{pmatrix}
1640\\158
\end{pmatrix}$}} &	{\footnotesize { $\begin{pmatrix}
1790\\ 396
\end{pmatrix}$}}&	{\footnotesize { $\begin{pmatrix}
2171\\ 445
\end{pmatrix}$}} &  \footnotesize{1.047}  \\[0ex]\hline\hline
\multicolumn{2}{|c|}{\multirow{4}{*}{Averages}}   & \footnotesize{3} &       &      &  &  &{\footnotesize { $\begin{pmatrix}
1512(25) \\ 161(36)
\end{pmatrix}$}} & {\footnotesize { $\begin{pmatrix}
1652(12) \\ 126(29)
\end{pmatrix}$}} &	\centering {\footnotesize { $\begin{pmatrix}
1986(373) \\ 471(301)
\end{pmatrix}$}} & -&  \\[0ex]
 &  & \footnotesize{4} &  &  &  & &{\footnotesize { $\begin{pmatrix}
1502(23) \\ 144(39)
\end{pmatrix}$}} & {\footnotesize { $\begin{pmatrix}
1649(13) \\ 117(37)
\end{pmatrix}$}} &	{\footnotesize { $\begin{pmatrix}
1745(80) \\ 220(95)
\end{pmatrix}$}} &	{\footnotesize { $\begin{pmatrix}
2191(241) \\ 392(301)
\end{pmatrix}$}}&   \\ [0ex] \hline \hline
\end{tabular}
\end{center}
\label{Table2}
\end{table*}
\clearpage

\clearpage

\begin{figure}[!h]
\begin{center}
\includegraphics[width=0.8\textwidth]{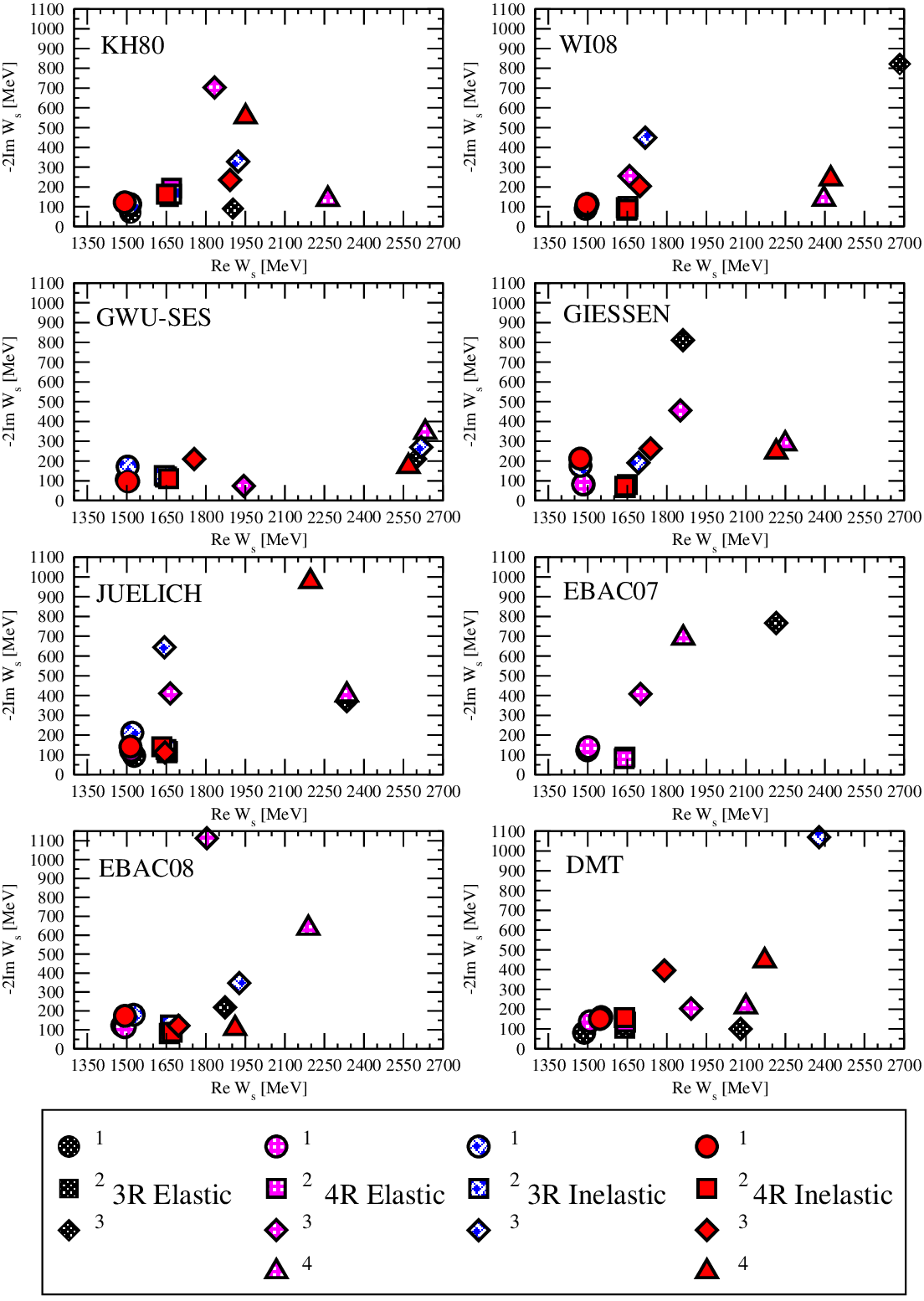}
\end{center} 
 \caption{Poles of a ``world collection" of PWA.}
 \label{Fig_poles_all}
\end{figure}

All obtained pole positions are shown in Fig.~\ref{Fig_poles_all}. \\
\vspace{0.65cm}

\clearpage
\subsection{ Individual comparison}
\subsubsection*{Preliminary considerations}
As it has been generally accepted, T-matrix pole positions are the most recommendable singularities to be compared with QCD. However, obtaining them  definitely means going 
into the complex energy plane while having at ones disposal only the physical T-matrix values (values for the real energy). This analytic continuation, however, has to be a 
model dependent procedure by definition, because there is no a priori rule  how to choose the analytic functional form which is to represent  a measurable subset out of all 
possible T-matrix values. Therefore, the reader has to be fully aware that the pole positions we find, and the pole positions given by the original publications have to be 
different \textbf{\textit{by definition}}, and the reason is that each investigated ``world collection" solution has its \textbf{\textit{own way}} how to analytically continue 
the measurable physical T-matrix values. However, comparing the number of needed poles, their distribution and genesis (genuine or dynamic) obtained  by our approach with those 
from original publication is certainly justified. It is also a convenient way to establish whether a certain pole is only a result of a poor knowledge of measured process, or 
indeed is a genuine singularity needed by the data, but still not yet well established. So, hereafter,  we analyze qualitative features of the partial wave singularity 
structure, and intentionally avoid to compare their numeric values. 
\subsubsection{KH80}
         The KH80 amplitudes are essentially single-channel partial wave data with some information about inelastic channels introduced through forward dispersion relations, 
and analyticity strictly imposed on the level of fitting procedure using Pietarinen expansion  \cite{Pietarinen}. As no assumption on the analytic functional form about  
partial-wave amplitudes has been done, search for resonance parameters is a separately defined procedure. Breit-Wigner parameters are obtained as a local fit in the resonance 
region with background contribution unitary added on the level of S matrices, and poles are extracted using single-channel pole position extraction methods (speed plot and 
Argand diagram). Original publication reported two poles. \\ 

In our approach we concur the existence of first two poles, and we find them strongly dominated by the elastic channel. \\ 

The third N(2090) pole is in our fit definitely needed. The fourth pole is allowed by our fit in both, single and coupled-channel constellation (improvement of the reduced 
$\chi_R^2$), but its quantitative constraint will need  more inelastic channels than only $\eta$N. In each configuration numerical values of third and fourth pole are not yet 
sufficiently well constrained. 

\subsubsection{GWU-SES and WI08}
 While the original publication gets the pole positions by analytically continuing energy dependent solution into the complex energy plane, an obvious advantage of our approach 
is that we can  obtain the pole positions independently from both, single energy (GWU-SES) and energy dependent (WI08) VPI/GWU solutions.  We have to remember that  VPI/GWU 
pole positions are extracted from the analytic form determined by their Chew-Mandelstam K-matrix approach, which is fitted \textit{directly to the data}, and not to their 
single channel solutions. Consequently,  the pole positions "corresponding" to their single energy solutions are \textit{by them} not yet discussed. In this paper we may use 
the same formalism for both, single energy and energy dependent solutions, and treat them as an independent input. Hence, we get two sets of solutions. \\

The general conclusion for both VPI/GWU solutions is the same, and it is very similar to the findings for the KH80 input: we confirm the existence of first two poles, and  find 
them strongly dominated by the elastic channel. The third N(2090) pole is in our fit definitely needed. The fourth pole is allowed by our fit in both, single and two channels 
constellation (improvement of the reduced $\chi_R^2$), but its quantitative constraint will need  more inelastic channels that $\eta$N. \\ 

It is very interesting to compare WI08 with GWU-SES.  In spite of the fact that the WI08 solution is seemingly very smooth above the second peak, definitely much smoother than 
the GWU-SES solution,  our model still requires the third and fourth pole almost in a same way for both solutions. The need for a third and fourth pole for the smooth WI08 
solution came as a surprise for us. Quantitatively, all pole positions are similar for both solutions: quite well defined for the first two poles, dominantly determined with 
the elastic channel. Inclusion of inelastic $\eta$N channel data modifies first two pole positions only slightly. However third and fourth pole positions remain strongly 
influenced.

\subsubsection{DMT amplitudes}
DMT collaboration has originally looked for the pole positions using the speed plot technique. They have established the existence of three poles, N(1535) , N(1650)  and a 
third pole corresponding to N(2090) - see ref. \cite{Che07}. However,  triggered by their old research of photo-production channels in which they had established the strong 
probability for the existence of  new S-wave resonant state in the vicinity of 1846 MeV  \cite{Che03,Yan03}, they have recently repeated the analysis and confirmed the 
existence of this new state at 1880 MeV \cite{Mainz2010}. \\

It is interesting to note that our procedure for DMT amplitudes  also indicates the existence of 4 poles. As seen in Fig. \ref{Fig_podaci_elastic}, our 3-resonant fits do miss 
some structure in elastic partial waves at higher energies requiring the increase in the number of parameters. Repeated fits with 4 resonances rectify this problem and at the 
same time show a significant improvement in the reduced  $\chi^2$.  So, our fits concur their latest findings  that the DMT S$_{11}$ solution really contains 4 poles  
\cite{Mainz2010}. 

\subsubsection{EBAC amplitudes}
  EBAC has produced three sets of partial wave amplitudes: the first, single channel set where only $\pi$N elastic data have been fitted - EBAC07 - \cite{Diaz07}, and two 
additional sets of amplitudes where data from more than one channel was used to constrain the fit; in this particular case $\pi$N and  $\eta$N channels. The  unpublished set 
\cite{Saghai2010} in a way supersedes the former 2008 analysis \cite{Dur08} where unpleasantly large change of $\pi$N elastic partial waves was needed to accommodate for the 
second channel. We have analyzed both sets of amplitudes wandering whether a significant change in poles between the two is found. However, as no numeric data for the 
unpublished set is available to us, we have attempted to "read off" the data directly from the graph, and that has introduced uncontrollable numeric instabilities. Therefore, 
we have decided to omit the EBAC10 preliminary data from our analysis until the final results are published.  \\

The EBAC group has in all three analyses used two bare poles, situated relatively high in energy (M $\geq$ 1.8 GeV), and reported two dressed poles corresponding roughly to 
N(1535) and N(1650). Third and fourth pole have not been found. 
Just as a preview, we can state that our analysis finds all three solutions very similar. For all three sets we confirm the existence of the first two poles, and they are 
strongly constrained by the $\pi$N elastic channel alone. However, our fits indicate that significant improvement reduced $\chi^2$ is achieved if the third and fourth poles are 
allowed for. These poles are needed by the fit, but still poorly determined by only two inelastic channels.  

\subsubsection{J\"{u}lich amplitudes}
Similarly to many, J\"{u}lich group fits their model to VPI/GWU data (to energy dependent WI08 set \cite{GWUWEB}), and very much like WI08, obtains a very smooth behavior above 
1800 MeV. The only difference with respect to WI08 is a different behavior of high energy tail: while the real part of J\"{u}lich amplitudes falls with energy and the imaginary 
part raises, in case of WI08 amplitudes the result is just opposite. Therefore, a difference between the two should not be found in cross section measurements, but only 
possibly in some polarization ones. They also report two S$_{11}$ poles. \\

Consequently, we expect that our results for pole positions of J\"{u}lich amplitudes show a very similar behavior to WI08, and that is fulfilled. \\

The most prominent feature of our analysis of WI08 amplitudes---that in spite of smooth high energy behavior we need more than two poles to fit the input---is confirmed for 
J\"{u}lich amplitudes as well. It is completely clear that we need at least three poles to satisfactorily reproduce the amplitude shape, and their amplitudes are in our 
analysis consistent with four S$_{11}$ poles. Very similar as before, the third and fourth poles are rather undetermined with only two channel constraints. We have discussed 
the possibility of finding extra poles in J\"{u}lich  amplitudes with M. D\"{o}ring in Zagreb last fall \cite{Duering2009}, and this possibility has not been entirely ruled out 
even in analytical continuation J\"{u}lich  method. They have simply not looked for the pole in that energy range.  However, even while  this pole might be around 1800 MeV, it 
must be rather far in the complex energy plane.  

\subsubsection{Giessen amplitudes}
  Giessen group also fits GWU-SES data in $\pi$N elastic channel, and gets a reasonable agreement with the input. The main difference with respect to ``world collection" 
amplitudes, again lies in the $\eta$N channel data. Nost results for this channel more or less agree within the N(1535) dominance range, but significantly deviate in the higher 
energy region. \\

The Giessen model assumes K-matrix Born approximation where the real part of the Green function is neglected, the analyticity is manifestly violated. Consequently, the 
comparison of poles obtained in our fit with poles of these amplitudes is more questionable, as the main assumption for the correct analytic continuation---that is the 
analyticity of the model---is not preserved for both models. \\

\subsection{Primary result: Averages}
As the main aim of the paper is to use one method in order to eliminate systematic uncertainties in pole extraction, we summarize our primary results. 
\subsubsection{$\pi$N  elastic channel only} 
\begin{figure}[!htb]
\begin{center}
\includegraphics[width=0.5\textwidth]{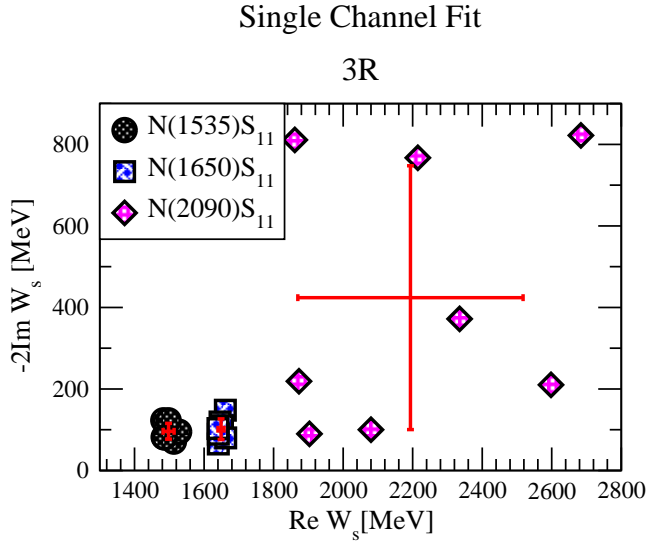}
\end{center} 
 \caption{ ``World collection" of poles for the three resonance single-channel fit.}
 \label{Fig_3R_elas}
\end{figure}

\begin{figure}[!h]
\begin{center}
\includegraphics[width=8.cm]{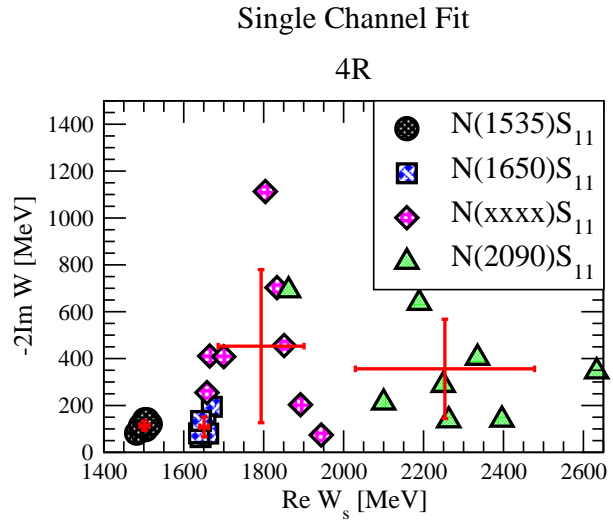}
\end{center} 
 \caption{ ``World collection" of poles for the four resonance single-channel fit.}
 \label{Fig_4R_elas}
\end{figure}

All pole positions and their averages are shown in Figs.~\ref{Fig_3R_elas} and \ref{Fig_4R_elas}. \\ \\ \noindent
\textit{\underline{Three resonant case}} \\ \\
As the  number of accepted S$_{11}$ resonances in  PDG \cite{PDG2010} is three, we first stooped our fit at 3 bare poles.
\vspace{0.2cm}

By inspecting 3R solutions in Table 1 and Fig. \ref{Fig_3R_elas}  we observe:
\begin{itemize}
   \item  First two poles N(1535) and N(1650)  are extremely well determined in all PWA.
   \item We find their average value to be 
\\ \\ $\overline{N(1535) \ S_{11}}$ \  = \ {\footnotesize { $\begin{pmatrix} 1498 \pm 16 \\ 96 \pm 20 \end{pmatrix}$}};
\\ \\  $\overline{N(1650) \ S_{11}}$ \  = \ {\footnotesize { $\begin{pmatrix} 1649 \pm 9 \\ 101 \pm 25 \end{pmatrix}$}}. 
    \item All PWA do need a third pole, but its position is extremely ill-defined;  KH80, Giessen and EBAC08 prefer the values between 1700 and 2000~MeV, while the rest have 
the values above 2000~MeV.  
    \item The resulting average value is poor \\ \\ 
	$\overline{N(2090) \ S_{11}}$ \  = \ {\footnotesize { $\begin{pmatrix} 2194 \pm 324 \\ 424 \pm 324 \end{pmatrix}$}}.			 
\end{itemize}
This separation in two preferred ranges of the third pole among different PWA permits us to speculate whether the fitting rules allow for the existence of the 4-th pole. \\ \\
\noindent
\textit{\underline{Four resonant case}} \\ \\
We have repeated the fit with 4-bare poles, and results are collected in Table 1. as 4R solutions. We show the result in Fig. \ref{Fig_4R_elas}. \\
\vspace{0.2cm}

By inspecting 4R solutions in Table 1 and Fig. \ref{Fig_4R_elas}  we observe:
\begin{itemize}
      \item We have found that all other PWA if not require, then are at least consistent with the 4 S$_{11}$ poles, even the EBAC amplitudes which are based on only 2 bare 
poles input.
    \item The reduced $\chi^2$ is either improved, or at least stays the same for all solutions; that justifies the inclusion of the fourth pole. 
  \item  First two poles N(1535)  and N(1650) are again very well determined in all PWA
   \item We find their average value to be 
    \\ \\ $\overline{N(1535) \ S_{11}}$ \  = \ {\footnotesize { $\begin{pmatrix} 1502 \pm 12 \\ 114 \pm 20 \end{pmatrix}$}}
   \\ \\ $\overline{N(1650) \ S_{11}}$ \  = \ {\footnotesize { $\begin{pmatrix} 1651 \pm 11 \\ 109 \pm 42 \end{pmatrix}$}}
    \item Contrary to our expectations, and in spite of the fact that the reduced $\chi^2$ is improved practically everywhere, the scatter in 3rd and 4th pole remain. 
    \item The resulting average value for third and fourth pole is poor 
	  \\ \\ $\overline{N(xxxx) \ S_{11}}$ \  = \ {\footnotesize { $\begin{pmatrix} 1793 \pm 108 \\ 453 \pm 327 \end{pmatrix}$}} 
   \\ \\ $\overline{N(2090) \ S_{11}}$ \  = \ {\footnotesize { $\begin{pmatrix} 2253 \pm 224 \\ 356 \pm 212 \end{pmatrix}$}}
\\ \\ The existence of the fourth pole is not convincing.          
\end{itemize}
 Due to the fact that third and fourth pole poorly couple to the elastic channel that is only used at this instant, we conclude that fitting other channels is inevitable if the 
improvement on the third and fourth pole parameters is to be achieved.   

\subsubsection{$\pi$N  elastic and $\pi$N $\rightarrow \eta$N data} 
\begin{figure}[!h]
\begin{center}
\includegraphics[width=8.cm]{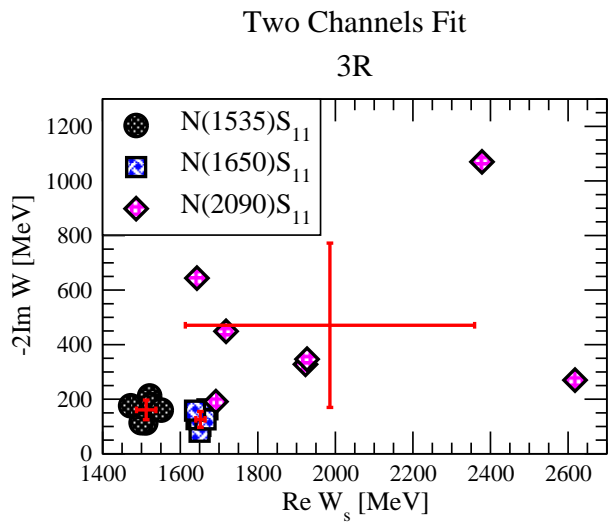}
\end{center} 
 \caption{ ``World collection" of poles for the three resonance, two channels fit.}
 \label{Fig_3R_inelas}
\end{figure}

\begin{figure}[!h]
\begin{center}
\includegraphics[width=8.cm]{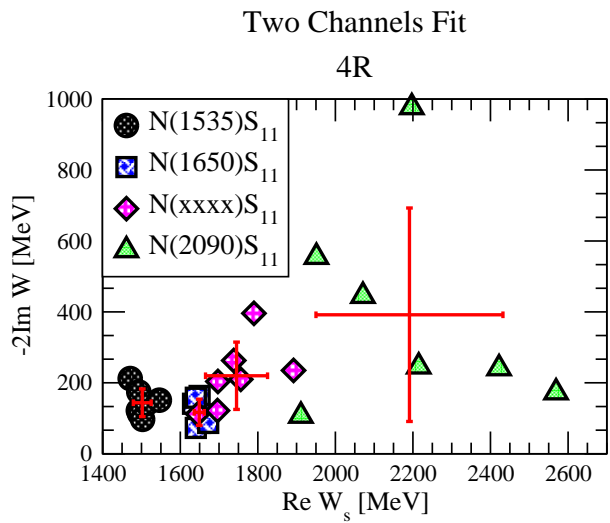}
\end{center} 
 \caption{ ``World collection" of poles for the four resonance, two channels fit.}
 \label{Fig_4R_inelas}
\end{figure}

The poor determination of third and fourth pole for the single channel fit confirms our former findings that inelastic channels are essential for fully constraining all 
resonant states (scattering matrix poles)  - see ref. \cite{Ceci06}. The problem with stability of minimization solutions lies in the fact that the $\eta$N channel data are 
old, scarce, and unreliable (for instance Brown data at higher energies - see discussion in ref. \cite{Bat98}), so  $\eta$N channel partial waves are imprecise. Even when being 
of lower quality, the  $\eta$N channel data still represent a valuable constraining condition, because the general trends of $\eta$N channel are to be simultaneously reproduced 
together with the details of elastic channel, and that is by no means simple.  The results of the fit are given in Table~2 and Figs.~\ref{Fig_3R_inelas} and 
\ref{Fig_4R_inelas}. \\ \\
\noindent
\textit{\underline{Three resonant case}} \\ \\
As the  number of accepted S$_{11}$ resonances in  PDG \cite{PDG2010} is three, we first stooped our fit at 3 bare poles. We show the result in Fig. \ref{Fig_3R_inelas}. \\
\vspace{0.2cm}

By inspecting 3R solutions in Table 2 and Fig. \ref{Fig_3R_inelas}  we observe: 
\begin{itemize}
   \item  First two poles N(1535)  and N(1650)  are extremely well determined in all PWA.
   \item We find their average value to be
    \\ \\  $\overline{N(1535) \ S_{11}}$ \  = \ {\footnotesize { $\begin{pmatrix} 1512 \pm 25 \\ 161 \pm 36 \end{pmatrix}$}}
   \\ \\ $\overline{N(1650) \ S_{11}}$ \  = \ {\footnotesize { $\begin{pmatrix} 1652 \pm 12 \\ 126 \pm 29 \end{pmatrix}$}}.

    \item If we compare these numbers with the result of single-channel, three resonance fit: 
  \\ \\ \textit{\footnotesize $\overline{N(1535) \ S_{11}}$} \  = \ {\footnotesize { $\begin{pmatrix} {\it 1498 \pm 16} \\ {\it 96 \pm 20} \end{pmatrix}$}}, 
  \\ \\ \textit{\footnotesize $\overline{N(1650) \ S_{11}}$} \  = \ {\footnotesize { $\begin{pmatrix} {\it 1649 \pm 9} \\ {\it 101 \pm 25} \end{pmatrix}$}},
 	\\ \\  we see that the difference is within one standard deviation. Real parts of the resonances are almost completely reproduced, while the imaginary parts are slightly  
shifted downwards.  
    \item All PWA do need a third pole, but it's position is again extremely ill-defined. 
    \item The resulting average value is poor 
		 \\ \\  $\overline{N(2090) \ S_{11}}$ \  = \ {\footnotesize { $\begin{pmatrix} 1986 \pm 373 \\ 471 \pm 301 \end{pmatrix}$}}.
 
\end{itemize}
\noindent
\textit{\underline{Four resonant case}} \\ \\
We have repeated the fit with 4-bare poles, and results are collected in Table 2. as 4R solutions. We show the result in Fig. \ref{Fig_4R_inelas}. 
\vspace{0.3cm}
 \\
By inspecting 4R solutions in Table 1 and Fig. \ref{Fig_4R_inelas}  we observe:
\begin{itemize}
   \item We have found that all PWA if not require, then are at least consistent with the 4 S$_{11}$ poles, even the EBAC amplitudes which are based on only 2 bare poles.
  \item The reduced $\chi^2$ is either improved, or stays the same for all solutions.  That justifies the inclusion of the fourth pole.
    \item  First two poles N(1535)  and N(1650) are again extremely well determined in all PWA
   \item We find their average value to be 
 \\ \\  $\overline{N(1535) \ S_{11}}$ \  = \ {\footnotesize { $\begin{pmatrix} 1502 \pm 23 \\ 144 \pm 39 \end{pmatrix}$}} 
 \\  \\  $\overline{N(1650) \ S_{11}}$ \  = \ {\footnotesize { $\begin{pmatrix} 1649 \pm 13 \\ 117 \pm 37 \end{pmatrix}$}}.
         \item The resulting average value for third and fourth pole are: 
		  \\ \\  $\overline{N(xxxx) \ S_{11}}$ \  = \ {\footnotesize { $\begin{pmatrix} 1745 \pm 80 \\ 220 \pm 95 \end{pmatrix}$}} 
         \\  \\  $\overline{N(2090) \ S_{11}}$ \  = \ {\footnotesize { $\begin{pmatrix} 2191 \pm 241 \\ 392 \pm 301 \end{pmatrix}$}}.
 
\item The scatter in the 3rd pole is significantly reduced, and the indications for its existence are strong. 
\item The existence of the fourth pole is strongly indicated, but still not quite convincing. 
\item If we compare these numbers with the result of single-channel, four resonance fit: \\ \\
	 \textit{\footnotesize $\overline{N(1535) \ S_{11}}$} \  = \ {\footnotesize { $\begin{pmatrix} {\it 1502 \pm 12} \\ {\it 114 \pm 20} \end{pmatrix}$}}, 
      \\ \\ \textit{\footnotesize $\overline{N(1650) \ S_{11}}$} \  = \ {\footnotesize { $\begin{pmatrix} {\it 1651 \pm 11} \\ {\it 109 \pm 42} \end{pmatrix}$}},
     \\ \\ \textit{\footnotesize $\overline{N(xxxx) \ S_{11}}$} \  = \ {\footnotesize { $\begin{pmatrix} {\it 1793 \pm 108} \\ {\it 453 \pm 325} \end{pmatrix}$}}, 
     \\ \\ \textit{\footnotesize $\overline{N(2090) \ S_{11}}$} \  = \ {\footnotesize { $\begin{pmatrix} {\it 2253 \pm 224} \\ {\it 356 \pm 212} \end{pmatrix}$}},

we conclude: \\  the $\eta$N channel data have confirmed the good constraint on N(1535) and N(1650) S$_{11}$ states,   they have improved the confidence limits for the 
existence of the new N(xxxx) S$_{11}$ state,  but they are definitely insufficient to constrain  the fourth S$_{11}$ pole.  \\  \\ Therefore, other channel partial waves have 
to be included.  
\end{itemize}

\clearpage
\section{Conclusions}

We have offered one model, the Zagreb realization of CMB model,  for extracting pole positions from a ``world collection" of partial-wave amplitudes which we treat as partial 
wave input data, and extracted the results. Using only one method enables us to make a statistical analysis of partial wave poles in a manner that we avoid the systematic error 
caused by the different assumptions on the partial wave analytic function form. We have in details explained the idea, and presented the results for the S$_{11}$ partial wave 
only.  \\  

We have analyzed the single channel fit (only one channel data are used to constrain the fit), and in details investigated what are the consequences of enlarging it to a 
two-channel one with the $\eta$N channel. We have concluded that even low quality data in the second channel are sufficient to notably constrain the arbitrariness of the poorly 
determined poles. However, we also concluded that for the third and fourth S-wave poles, $\eta$N channel is not sufficient. \\

We found that the first two S$_{11}$ poles are extremely well defined by elastic channel and that the included inelastic $\eta$N channel introduces only small modifications of 
the elastic channel result. \\

We have shown that all members of the partial wave ``world collection", in spite of the fact that some of them have assumed only two S-wave resonant states, are consistent with 
at least three T-matrix poles. We have also demonstrated that there is a strong statistical indications that the fourth pole is present in each of the ``world collection" 
member, despite the fact that no one has seen it up to now. The exception is the latest DMT analyses \cite{Mainz2010} where all four poles are detected. The new S$_{11}$ pole 
we have found can be identified with the S$_{11}$(1846) pole  seen in photo-production channel  by the DMT collaboration \cite{Che03,Yan03}.  \\

We affirm that the results of the 4-resonant, double channel fit should be treated as a final result, and we offer the world average: 
\begin{equation}
  \begin{aligned}
& {\rm N}(1535) \ {\rm S}_{11}= \footnotesize{\begin{pmatrix} 1502 \pm 23 \\ 144 \pm 39 \end{pmatrix}} \\
& {\rm N}(1650) \ {\rm S}_{11}= \footnotesize{\begin{pmatrix} 1649 \pm 13 \\ 117 \pm 37 \end{pmatrix}} \\
& {\rm N}(xxxx) \ {\rm S}_{11}= \footnotesize{\begin{pmatrix} 1745 \pm 80 \\ 220 \pm 95 \end{pmatrix}} \\
& {\rm N}(2090) \ {\rm S}_{11}= \footnotesize{\begin{pmatrix} 2191 \pm 241 \\ 392 \pm 301 \end{pmatrix}} \\
 \end{aligned}
\end{equation}


\clearpage

 
\begin{thebibliography}{12} 
\bibitem{Dal70} R.H. Dalitz and R.G. Moorhouse, Proc. Roy. Soc. Lond. A318 (1970) 279-298.
\bibitem{Man58} S. Mandelstam,  Phys. Rev. {\bf 112} (1958) 1344. 
\bibitem{Mar70} A.D. Martin, T.D. Spearman, \emph{Elementary particle theory}, North-Holland Publishing Company, Amsterdam, 1970. 
\bibitem{Arndt1}  R. A. Arndt {\it et all.}, Phys. Rev. {\bf C 74} (2006) 045205. 
\bibitem{Manley1}   D. M. Manley and E. M. Salesky, Phys. Rev {\bf D 45} (1992) 4002.
\bibitem{Manley2}   M. Manley, Phys. Rev. {\bf D 51} (1995) 4837.
\bibitem{Cutcosky1}   R. E. Cutkosky {\it et all.}, Phys. Rev. {\bf D 20} (1979) 2804.
\bibitem{Bat1}   M. Batini\'{c}, {\it et al.}, Phys. Rev. {\bf C 51}, 2310 (1995);
M. Batini\'{c}, {\it et al.}, Physica Scripta {\bf 58}, 15, (1998).
\bibitem{TPVrana}   T. P. Vrana, S. A. Dytman, T. -S. H. Lee, Phys. Rep. {\bf 328} (2000) 181.
\bibitem{Fla76} S.M. Flatt\'{e}, Phys. Lett. {\bf B 63}, 224 (1976);   S. M. Flatt\'{e} {\it et al.}, Phys. Lett. {\bf B 38}, 232 (1972).  
\bibitem{Pole_vs_BW} N. G. Kelkar, M. Nowakowski, K. P. Khemchandani, Sudhir R. Jain,  Nucl.Phys.\textbf{ A 730} (2004) 121-140.
\bibitem{KH80} G. H\"{o}hler, in {\em Pion-Nucleon Scattering}, Landolt-B\"{o}rnstein, Vol {\bf I/9b2} (Springer-Verlag, Berlin, 1983);   G. H\"{o}hler, A. Schulte, $\pi N$ 
Newsletter, {\bf 7} (1992) 407.
\bibitem{PDG1998} G.~H\"ohler, {\it Against Breit-Wigner parameters -- a pole-emic},
  in C.~Caso {\it et al.}  [Particle Data Group],
  Eur.\ Phys.\ J.\  C {\bf 3}, 624 (1998).
\bibitem{PDG2000} G. H\"{o}hler, {\it RESULTS ON $\Delta$(1232) RESONANCE PARAMETERS:
A NEW $\pi$N PARTIAL WAVE ANALYSIS}, in NSTAR2001, Proceedings of the Workshop on the Physics of Excited Nucleons, Edts. D. Drechsel, L. Tiator, World Scientific Publishing Co. 
(2001), Pg.185.
\bibitem{PDG2010} K. Nakamura et al. (Particle Data Group), J. Phys. \textbf{G 37}, 075021 (2010).
\bibitem{Ceci08} S, Ceci, J. Stahov, A. \v{S}varc, S. Watson  and B. Zauner, Phys. Rev. \textbf{D 77}, 116007 (2008).
\bibitem{Mainz2010} L. Tiator, S.S. Kamalov, S. Ceci, G. Y. Chen, D. Drechsel, A. \v{S}varc and S. N. Yang, Phys. Rev. \textbf{C 82}, 055203 (2010).
\bibitem{Eisenbud1} {  L. Eisenbud, disertation, Princeton, June 1948 (unpublished)}
\bibitem{Wigner1} {  E. P. Wigner, Phys. Rev. {\bf 98}, 145 (1955)}
\bibitem{Wigner2} {  E. P. Wigner, L. Eisenbud Phys. Rev. {\bf 72}, 29 (1947)}
\bibitem{Bohm} {  D. Bohm, {\itshape Quantum Theory} (Prentice-Hall, New York, 1951)}
\bibitem{Suzuki1} {  N. Suszuki, T. Sato, T. -S. H. Lee,  Proceedings of the Menu2007 11th International Conference on Meson-nucleon Physics and the structure of the Nucleon, 
J\"{u}elich 2007, edited by H. Machner, S Krewald, eConf C070910 (2007) 407}
\bibitem{Chew60} G. F. Chew and S. Mandelstam, Phys. Rev. {\bf 119},  467 (1960).
\bibitem{Olle99} J. A. Oller and E. Oset,Phys. Rev. {\bf D 60}, 074023 (1999).
\bibitem{Ani06} V.V. Anisovich, International Journal of Modern Physics,  {\bf A 21}, 3615 (2006).
\bibitem{Arn04}R. A. Arndt, W.J. Briscoe, I.I. Strakovsky, R.L. Workman, and M.M. Pavan, Phys. Rev. {\bf C69}, 035213 (2004).
\bibitem{EBAC} N. Suzuki, T. Sato and T. -S. H, Lee, Phys. Rev. \textbf{C 79}, 025205 (2009).
\bibitem{Feu98} T. Feuster and U. Mosel1, Phys. Rev. \textbf{C 58}, 457 (1998).
\bibitem{Che03} Guan-Yeu Chen, Sabit Kamalov, Shin Nan Yang, Dieter Drechsel, Lothar Tiator, Nuclear Physics \textbf{A 723},  447  (2003).
\bibitem{Che07} Guan Yeu Chen, S. S. Kamalov, Shin Nan Yang, D. Drechsel and L. Tiator, Phys. Rev. \textbf{C76}, 035206 (2007).
\bibitem{Giessen} V. Shklyar, H. Lenske, U. Mosel, Phys. Rev. \textbf{C 72},  015210 (2005), and private communication.
\bibitem{Yan03} S. N. Yang, G.-Y. Chen, S. S. Kamalov, D. Drechsel and L. Tiator, Nucl. Phys. \textbf{A 721} 401c (2003); S. N. Yang, G.-Y. Chen, S. S. Kamalov, D. Drechsel and 
L. Tiator, Int. Journal of Modern Physics, \textbf{A 20} 1656 (2005).
\bibitem{Hedim2010} H. Osmanovi\'{c}, S. Ceci, A. \v{S}varc, M. Had\v{z}imehmedovi\'{c} and J. Stahov, submitted to Phys. Rev. C. 
\bibitem{Bat98}  M. Batini\'{c}, {\it et al.}, Phys. Rev. {\bf C 51}, 2310 (1995);
M. Batini\'{c}, {\it et al.}, Physica Scripta {\bf 58}, 15, (1998).
\bibitem{Cut79} R. E. Cutkosky {\it et al.}, Phys. Rev. {\bf D 20}, 2839 (1979).
\bibitem{GWUWEB} http://gwdac.phys.gwu.edu/analysis/pin\_analysis.html.
\bibitem{Diaz07} B. Juli\'{a}-D\'{i}az, T.-S. H. Lee,1 A. Matsuyama, and T. Sato, Phys. Rev. \textbf{C 76}, 065201 (2007).
\bibitem{Dur08} J. Durand, B. Juli\'{a}-D\'{i}az, T.-S. H. Lee, B. Saghai, and T. Sato, Rev. \textbf{ C 78}, 025204 (2008).
\bibitem{Juelich} M. D\"{o}ring, C. Hanhart,, F. Huang, S. Krewald, U.-G. Meissner, Nucl.Phys,  \textbf{A 829}, 170 (2009); C. Sch\"{u}tz, J. Haidenbauer, J. Speth and J. W. 
Durso, Phys. Rev. \textbf{C 57} 1464 (1998); O. Krehl, C. Hanhart, S. Krewald and J. Speth, Phys. Rev. \textbf{C 62} 025207 (2000); A. M. Gasparyan, J. Haidenbauer, C. Hanhart 
and J. Speth, Phys. Rev. \textbf{C 68} 045207 (2003).
\bibitem{Ceci06} S. Ceci, A. \v{S}varc, and B. Zauner, Phys. Rev. Lett. \textbf{97}, 062002 (2006).
\bibitem{Duering2009} M. D\"{o}ring, B. Diaz private communications.
\bibitem{Pietarinen} E. Pietarinen, Nuovo Cimento, {\bf 12A}, 522 (1972).
\bibitem{Saghai2010}   B. Saghai, et al, EBAC meeting on "Extraction of nucleon resonances", May 24 - 26, 2010, JLab, Virginia, USA; \\
\mbox{http://ebac-theory.jlab.org/workshop\_meeting/m2010/talks/Saghai-ebac-m2010.pdf}
\end{thebibliography}
\end{document}